\documentclass[twocolumn,amsmath,amssymb,10pt,superscriptaddress,a4paper,letterpaper,fleqn]{revtex4-1}
\usepackage{amssymb}
\usepackage{epsfig}
\usepackage{graphicx}
\usepackage{dcolumn}
\usepackage{array}
\usepackage{bm}
\usepackage{fancyheadings}
\usepackage{longtable}
\usepackage{multirow}
\usepackage{float}
\usepackage{afterpage}  
\usepackage{color}

\setlongtables
\usepackage[breaklinks=true,linkbordercolor={1 1 1}]{hyperref}

\begin{document}
\setcounter{page}{1}

\title{
\qquad \\ \qquad \\ \qquad \\  \qquad \\  \qquad \\ \qquad \\ 
MCNP6 Study of Fragmentation Products \\
from $^{112}$Sn+ $^{112}$Sn and  $^{124}$Sn+ $^{124}$Sn
at 1 GeV/nucleon
}

\author{S.G. Mashnik}
\email[Corresponding author, electronic address:\\ ]{mashnik@lanl.gov}

\author{A.J. Sierk}
\affiliation{Los Alamos National Laboratory, Los Alamos, NM 87545, USA }

\date{\today}

\begin{abstract}
Isotope production cross sections from 
 $^{112}$Sn + $^{112}$Sn and  $^{124}$Sn + $^{124}$Sn
reactions at 1 GeV/nucleon, which were measured recently at GSI using the
heavy-ion accelerator SIS18 and the Fragment Separator (FRS),
have been analyzed with the latest 
Los Alamos Monte-Carlo transport code MCNP6
using the LAQGSM03.03 event generator.
MCNP6 reproduces reasonably well all the measured cross sections.
Comparison of the MCNP6 results with the measured data
and with calculations by a modification of the Los Alamos
version of the Quark-Gluon String Model allowing for 
multifragmentation processes
in the framework of the Statistical Multifragmentation Model (SMM)
by Botvina and coauthors, as realized in the
code LAQGSM03.S1, does not suggest unambiguous
evidence of a multifragmentation signature.
\end{abstract}
\maketitle


\lhead{ND 2013 Article $\dots$}
\chead{NUCLEAR DATA SHEETS}
\rhead{A. Author1 \textit{et al.}}
\lfoot{}
\rfoot{}
\renewcommand{\footrulewidth}{0.4pt}

\section{ INTRODUCTION}
MCNP6 \cite{MCNP6}
is used in various applications involving reactions 
induced by neutrons and other particles, but also  may be
applied to heavy-ion collisions at relativistic energies.
The Los Alamos version of the Quark-Gluon String Model (LAQGSM),
implemented in the code LAQGSM03.03 \cite{Trieste08},
is the main ``workhorse''
(event generator) used by MCNP6 to describe relativistic heavy-ion
interactions.  It is critical that it be able to describe such reactions
as well as possible; therefore, it is extensively
validated and verified against available experimental data
and calculations by other models
(see, e.g., \cite{EPJP2011} and references therein). 
So far, for relativistic heavy-ion collisions,
MCNP6 has been compared mostly with different particle spectra
measured from various reactions \cite{EPJP2011}
and much less with data on isotope-production yields.
To remedy this lack, we test MCNP6 and the LAQGSM03.03 \cite{Trieste08}
and LAQGSM03.S1 \cite{Laq03.S1.G1} versions of LAQGSM against the
recent GSI measurements of fragmentation products in the
reactions $^{112}$Sn + $^{112}$Sn and  $^{124}$Sn + $^{124}$Sn
at 1 GeV/A \cite{Fohr2011}.
These data are interesting because we may use them to study
the influence of the isotopic composition of the projectile and
target on the kinematical properties of projectile 
residues, which may help improve understanding of
the physics of nuclear-fragmentation reactions.

\begin{figure}[!h]
\vspace*{2mm}
\includegraphics[width=1.0\columnwidth]{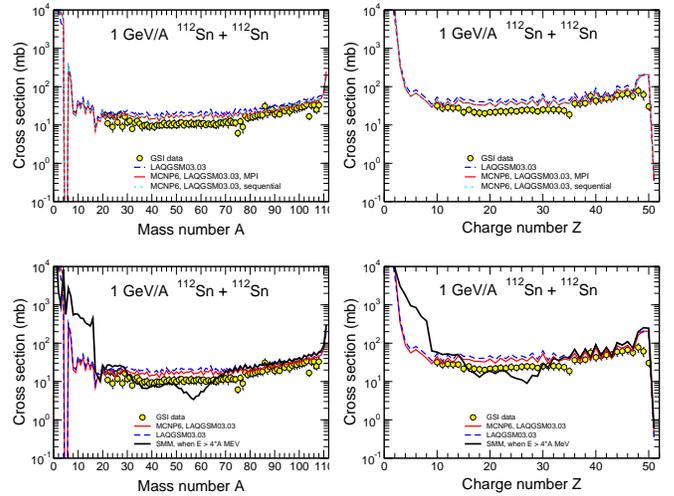}
\caption{Experimental \cite{Fohr2011} mass and charge distributions 
of products from 1 GeV/A $^{112}$Sn + $^{112}$Sn (yellow points) compared
with results by LAQGSM03.03 used as a stand-alone code (blue dashed lines)
and with MCNP6 calculations in parallel (MPI, red solid lines) and
sequential (green dashed lines) modes using the LAQGSM03.03 event generator
are in the upper plots, while results from the LAQGSM03.S1 version of
LAQGSM, which takes into account multifragmentation processes with
SMM \cite{SMM}, for nuclei with excitation energies $E$ above 
$4 \times A$ MeV (black solid lines), are in the lower plots.
}
\label{fig1}
\end{figure}

\begin{figure*}[!htb]
\includegraphics[width=1.0\textwidth]{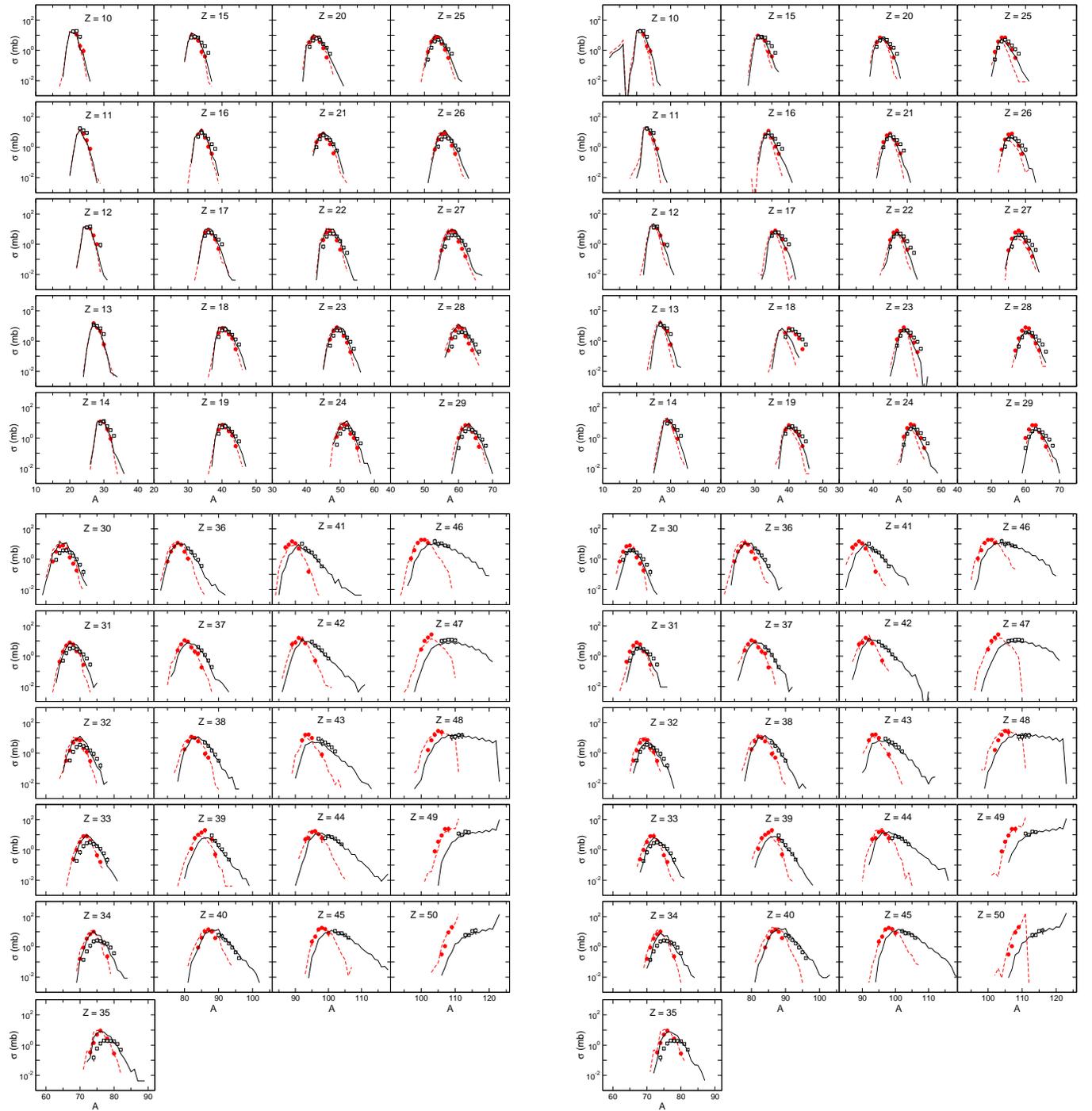}
\caption{Isotopic cross sections of all measured fragments
\cite{Fohr2011} in the reactions 1 GeV/A $^{112}$Sn+$^{112}$Sn
(red solid circles) and $^{124}$Sn+$^{124}$Sn (black open squares)
compared with results from LAQGSM03.03 \cite{Trieste08} (red dashed
and black solid lines, correspondingly) in the left panel and those
from LAQGSM03.S1 \cite{Laq03.S1.G1}, which accounts for multifragmentation 
processes with SMM \cite{SMM} for nuclei with excitation
energies $E \ge 4 \times A$ MeV, in the right panel.
}
\label{fig2}
\end{figure*}

\section{Results}

As part of the MCNP6 Verification and Validation (V\&V) process, we
calculate both $^{112}$Sn+$^{112}$Sn and $^{124}$Sn+$^{124}$Sn reactions
using the LAQGSM03.03 event generator, both running MCNP6 in a sequential
mode and in parallel using MPI.
We also calculate both reactions with LAQGSM03.03 used as a stand-alone 
code, outside MCNP6. As expected, all results obtained with MCNP6
run with MPI coincide with the ones calculated in a sequential mode.
Similarly, all the results obtained with MCNP6 using LAQGSM03.03
are practically the same
as the calculations done with LAQGSM03.03 by itself,
with only a tiny uniform difference in the absolute values of all
cross sections, due to different values for the total reaction
cross-section normalizations used by the two code systems.
In principle, there could be other problems with the implementation of
LAQGSM into MCNP6, but there are no apparent discrepancies for these
particular reactions.  Examples of some results from our study for
the  $^{112}$Sn+$^{112}$Sn reaction, are presented in the upper 
plots of Fig.~1.

The LAQGSM03.03 version of LAQGSM, as implemented in MCNP6, does not
account for multifragmentation of highly excited nuclei. It models
Fermi Break-up of light nuclei with $A < 13$ (see details in \cite{Trieste08}), 
but this is not what one usually means by ``multifragmentation''
(see, e.g., \cite{SMM} and references therein). In addition, Fermi
Break-up is only used in LAQGSM for nuclei with $A < 13$.

To investigate further the mechanisms of nuclide production in these
reactions, we calculate both reactions with a version of LAQGSM,
LAQGSM03.S1 \cite{Laq03.S1.G1},
which accounts for multifragmentation processes using the Statistical
Multifragmentation Model (SMM) of Botvina et al. \cite{SMM}.
One of the most important details in SMM
is the condition for the  transition between the multifragmentation 
and vaporization (or evaporation) modes of a reaction, which is 
determined by the temperature
of the excited nucleus, or by its excitation energy, $E$. 
It is believed that such a transition should occur at excitation energies
$E/A$ from $\sim 3$ to $\sim 5$ MeV/nucleon, depending on the concrete
type of the reaction (see the recent review by Borderie and Rivet \cite{Borderie08}
and references therein). We perform calculations with LAQGSM03.S1 for the
transition condition defined by
$E/A$ equal to 
4 MeV/nucleon.

We find that for these reactions
the results obtained with LAQGSM03.S1 are close to the values
of the product cross-section yields calculated with LAQGSM03.03 
or MCNP6. Examples of mass and charge distributions 
of products from the $^{112}$Sn+$^{112}$Sn
reaction calculated with LAQGSM03.S1
are compared with similar 
results by LAQGSM03.03 and by MCNP6 with LAQGSM03.03 
in the lower plots of Fig.~1.

A more detailed comparison of results by LAQGSM03.S1
and LAQGSM03.03 in MCNP6 is presented in Fig.~2,
where the left panel shows isotopic cross sections of all
measured products from both reactions compared with
calculations by LAQGSM03.03, while the right panel shows the
same data compared with results by LAQGSM03.S1.
A careful one-by-one comparison of the plots from the left
panel with the ones from the right panel does not suggest
an unambiguous signature of multifragmentation reactions
in this data.  We came to a similar conclusion
in Ref. \cite{SMMvsEvap} while analyzing other comparable
heavy-ion reactions. Mancusi et al. also made
a similar conclusion from a study of 1-GeV
proton-nucleus reactions \cite{Mancusi2011}.

\vspace*{5mm}
\section{ CONCLUSIONS}
The recent GSI measurements
$^{112}$Sn + $^{112}$Sn and  $^{124}$Sn + $^{124}$Sn
at 1 GeV/nucleon have been analyzed with the
Los Alamos transport code MCNP6,
using the LAQGSM03.03 event generator.
MCNP6 reproduces reasonably well all the measured cross sections.
All the MCNP6 results obtained in a sequential run
coincide with those obtained using a parallel version.
Comparisons of our MCNP6 results with the measured data do not
differ by large amounts from those from calculations by a modified 
LAQGSM allowing for multifragmentation processes
in the framework of the Statistical Multifragmentation Model (SMM).
Within the constraints of these particular models, there does not
seem to be any evidence unambiguously suggesting a multifragmentation
signature in these reactions.

\end{document}